\documentstyle[12pt]{article}
\def\be{\begin{equation}}
\def\ee{\end{equation}}
\begin{document}

\title{Superfluidity in the stochastic limit}
\author{L.Accardi, S.V.Kozyrev}
\maketitle

\centerline{\it Centro Vito Volterra, Universita di Roma Tor Vergata}

\begin{abstract}
In the present paper we outline the stochastic limit approach
to superfluidity.
The Hamiltonian describing the interaction between the Bose
condensate and the normal phase is introduced.
Sufficient in the stochastic limit condition of superfluidity is proposed.
Existence of superfluidity in the stochastic limit of this system is proved
and the non--linear (quadratic)  equation of motion
describing the superfluid liquid is obtained.
\end{abstract}

\section{Introduction}

The theory of superfluidity was developed by L.D.Landau and N.N.Bogoliubov,
cf. \cite{Landau3}, \cite{Landau}, \cite{B2}, \cite{B},
for an introduction see \cite{BB}. An analogue of the approach of
\cite{B2}, \cite{B} was applied to superconductivity theory,
cf. \cite{BTS}.

The essence of the superfluidity phenomenon is that the Bose
condensate becomes superfluid and friction between condensate
and normal phase disappears. N.N.Bogoliubov in
\cite{B2}, \cite{B} found that this can be explained as an effect
of stabilization of the condensate by interaction between particles.

In the present paper, using as a starting point the approach of
\cite{B2}, \cite{B}, we introduce an Hamiltonian that describes
the interaction between the Bose condensate and the normal phase and
investigate this Hamiltonian using the stochastic limit
approach cf. \cite{book}.

The name stochastic limit is due to the property that
in this approach the quantum field is approximated by
a quantum white noise and the Schr\"o\-din\-ger equation
is approximated by a white noise Hamiltonian equation.
In the stochastic limit we start from an Hamiltonian of the form
$$
H=H_0+\lambda H_I
$$
where $\lambda$ is a coupling constant and we make the time rescaling
$t\to t/\lambda^2$ in the solution of the Schr\"o\-din\-ger equation in
interaction picture $U^{( \lambda )}_t=e^{itH_0} e^{-itH}$,
associated to the Hamiltonian $H$, i.e.
$$
{ \frac{\partial}{ \partial t} }  U^{( \lambda )}_t
=-i { \lambda  } H_I (t) \ U_t^{
(\lambda )}\qquad , \ U_0^{(\lambda )} = 1
$$
with $H_I(t)=e^{it H_0}H_Ie^{-itH_0}$ (the {\it evolved interaction
Hamiltonian}).
This leads to the rescaled equation
$$
{ \frac{\partial}{  \partial t} }  U^{( \lambda )}_{t/\lambda^2}
=- {\frac{i}{\lambda}} H_I (t/\lambda^2) \
U_{t/\lambda^2}^{(\lambda )}
$$
and one wants to study the limits, in a topology to be specified,
$$
\lim_{\lambda\to0}H_I({t/\lambda^2})= H_t
$$
$$
\lim_{\lambda\to0}U^{(\lambda)}_{t/\lambda^2}= U_t
$$
Moreover one wants to prove that $U_t$ is the solution of the
white noise Hamiltonian equation
$$
\partial_t U_t\,=-iH_tU_t\quad;\qquad U_0=1
$$
which is equvalent to a quantum stochastic differential equation.

The structure of the present paper is as follows.
In section 2 we discuss the Bogoliubov--Landau condition
of superfluidity.
In sections 3 and 4 we review some basic facts from \cite{BB} on Bose
condensation in ideal and non--ideal Bose gases.
In section 5 we propose the Hamiltonian of condensate--normal phase
interaction and investigate this model using the stochastic limit.
In section 6 we construct by the stochastic golden rule the master equation
for the condensate--normal phase interaction and prove
the existence of superfluidity in this system.
We prove that the standard Bogoliubov--Landau condition is not
sufficient to garantee the superfluidity in the stochastic limit
and introduce a procedure to overcome this problem.
In section 7 we sum up the main conclusion that can be draw from our results.

\section{The Landau and Bogoliubov ideas on superfluidity theory}

Tissa and London conjectured that the existence of the Bose condensate
can explain superfluidity in the sense that the condensate
corresponds to the superfluid component and the temperature phase
corresponds to normal component.

By Landau's argument \cite{Landau3}, \cite{Landau}
the existence of a condensate itself is not enough to prove
superfluidity and moreover there is no superfluidity in ideal Bose gas.

Consider the condensate moving with velocity $u$ and the non--moving
normal state. Suppose that the friction of the condensate with
the normal phase stimulates the transition of a particle in condensate from
the state with velocity $u$ to the state with velocity
$u-{p\over m}$.
Then this particle will have energy ${(mu-p)^2\over 2m}$.
Before the transition the particle had energy ${mu^2\over 2}$.
Since the transition is due to friction with normal state
this corresponds to the excitation of
the particle of the normal phase with energy
${p^2\over 2m}$. We get the energy difference
$$
\hbox{ Energy after transition + Energy of excitation --
Energy before transition}=
$$
$$
={(mu-p)^2\over 2m} + {p^2\over 2m}-{mu^2\over 2}=
{p^2\over m} -pu<0 \quad\hbox{  for small } p
$$
This means that such transitions are energetically possible and
that the condensate is non--stable.

Bogoliubov found \cite{B2}, \cite{B}
that the interaction between the particles in thecondensate
may stabilize the condensate and modify the energy difference in such
a way that transitions to slower motion due to friction will
proceed with positive energy difference and therefore will
be energetically forbidden.

Bogoliubov found that interaction makes the state of normal phase
instable. To stabilize the state one has to make a Bogoliubov
transformation. This procedure changes the dispersion $E(k)$
of excitations of normal phase, cf.
\cite{B2}, \cite{B}, \cite{BB}, \cite{BTS}.

After the canonical transformation the energy difference
for transition $p\mapsto p-k$ is given by
\be\label{energydiff}
E(k)+\varepsilon(p-k)-\varepsilon(p)
\ee
where $\varepsilon(p)={p^2\over 2m}$ and $E(p)$
has different dependence on $p$.
In this case one can overcome the Landau objection.
Consider the following examples.

\bigskip

\noindent
EXAMPLE 1:\quad
Consider radiative dispersion $E(k)=c|k|+\dots$.
In this case $E(k)$ for small enough $k$ is proportional to $|k|$ and
(\ref{energydiff}) takes the form
\be\label{example1}
c|k|-pk +\frac{1}{2}\,k^2>0,\qquad |p| < c
\ee

\noindent
EXAMPLE 2:\quad
Consider the polaron model with $E(k)=\omega$.
In this case (\ref{energydiff}) becomes
$$
\omega-kp+\frac{1}{2}k^2>0,\qquad |p|< \sqrt{\omega}
$$

Let us show that the results of Landau and Bogoliubov discussed above
are connected with the stochastic limit approach.
One of important properties of the stochastic limit is that
the main properties of the dynamics is controlled by
$\delta$--functions of energy differences of the form
\be\label{deltaenergydiff}
\delta(E(k)+\varepsilon(p-k)-\varepsilon(p))
\ee
(this coincides with the $\delta$ of (\ref{energydiff})).
The possibility to control the dynamics by exploiting
such $\delta$--functions
was called in \cite{spinboson} the Cheshire Cat effect.
We find that the Bogoliubov--Landau condition of superfluidity
is connected with the Cheshire Cat effect developed
in the stochastic limit approach.

\section{Condensation of an ideal Bose gas}

In the present section we discuss standard material on
Bose condensation of an ideal Bose gas, cf. \cite{BB}.
The Hamiltonian of an ideal Bose gas in second quantization
is given by
$$
H=\int \omega(p) b^*(p) b(p) dp
$$
The grand canonical ensemble Hamiltonian is
$$
\Gamma=H-\mu N,\qquad N=\int  b^*(p) b(p) dp
$$
The equilibrium state of the gas is the unique mean zero gauge invariant
state with expectations
$$
\langle b^*(p) b(p)\rangle =\langle n(p) \rangle =
{1\over e^{\beta(\omega(p)-\mu)}-1}
$$
giving the density of particles with momentum $p$ and where
$\beta={1\over\theta}$ is the inverse temperature. We take $\mu\le 0$
since $\langle n(p) \rangle\ge 0$.

The integral over the density of degrees of freedom of gas is equal to
the density of the gas:
$$
\int \langle n(p) \rangle dp=\rho
$$

To describe Bose condensation, we consider the zero mode of the field
$b_0$ as singled out from the other modes
$$
[b_0,b^*_0]=1
$$
and take, as the reference state of the field, the state
not the equilibrium one but a gaussian state of the form
$$
\langle n(p) \rangle =
c\delta(p)+{1\over e^{\beta\omega(p)}-1}
$$
Notice that the chemical potential in this formula is taken
equal to zero.

We consider the constant $c=c(\theta)$ depending on the temperature.
To determine this dependence we use the conservation of number of particles,
given by the following:
\begin{equation}\label{conservation}
c+\int{1\over e^{\beta\omega(p)}-1} dp=\rho
\end{equation}
Therefore
$$
c=\rho\left(1-\left(\theta\over \theta_c\right)^{3\over2}\right)
$$
where $d=3$ and $\omega(p)={p^2\over 2m}$, $\theta\le \theta_c$.
The critical temperature
$\theta_c$ is the temperature when the integral in (\ref{conservation})
becomes equal to the total density of particles $\rho$ and condensate
disappears.

This implies the state of the Bose gas in the form
$$
\langle n(p) \rangle =
\rho\left(1-\left(\theta\over \theta_c\right)^{3\over2}\right)\delta(p)+
{1\over e^{\beta\omega(p)}-1}
$$

\section{Condensation of a non--ideal Bose gas}

In the present section we review material from \cite{BB}
containing the discussion of the Bose condensation of
non--ideal Bose gas and prove the applicability of
the Bogoliubov--Landau condition of stability of the condensate.

The second quantized Hamiltonian of a non--ideal Bose gas
(for the grand canonical ensemble) is:
$$
\Gamma=\sum_p(\omega(p)-\mu) b^*(p) b(p)  +
{\lambda\over2V}\sum_{p_1+p_2=p'_1+p'_2}
g(p_1-p'_1) b^*({p_1}) b^*({p_2})b(p'_2) b({p'_1})
$$
We consider a state such that almost all particles with $p=0$
are in the condensate. This implies
$$
\langle b^*_0 b_0 \rangle=N_0,\qquad N_0>>1
$$
where we denote $b^*_0$ the creation of the particle with $p=0$.
Since
$$
\langle b_0 b^*_0 \rangle=N_0+1
$$
and $N_0+1$ is almost equal to $N_0$, we can consider $b_0$ as a classical
(commuting) variable with
$$
|b_0|^2=b^*_0 b_0=b_0^* b_0=N_0,\qquad b_0=\sqrt{N_0}e^{i\phi}
$$
This allows to simplify the Hamiltonian $\Gamma$. Since $b(p)$ with $p\ne 0$
are not in the condensate, amplitudes of $b(p)$ are small with respect to
$b_0$. Therefore we can keep in the Hamiltonan $\Gamma$ only the
terms of second order in $b(p)$, $p\ne 0$
and skip higher order terms (3 and 4).

This mean field approximation will be valid only when almost all particles
are in the condensate, i.e. for a temperature close to zero:
$\theta=0$.

In this approximation the fourth order term in $\Gamma$
takes the form
$$
{\lambda\over 2V}\biggl(g(0)N^2_0+
b_0^{*2}\sum_p g(p) b(p) b({-p}) +
b_0^{*}b_0 \sum_p (g({p})+g({-p})) b(p) b({p})+
$$
$$
+2b_0^{*}b_0 \sum_p g({0}) b(p) b({p})+
b_0^{2}\sum_p g(p) b^*(p) b^*({-p})
\biggr)
$$
Using the conditions $g(p)=g(-p)$, $\omega(0)=0$ we get for $\Gamma$
$$
\Gamma={\lambda N_0^2\over 2V}g(0)-\mu N_0+
\sum_{p\ne 0}\left(\omega(p)-\mu +{\lambda N_0\over V}g(0)
+{\lambda N_0\over V}g(p) \right)b^*(p) b(p) +
$$
$$
+{\lambda \over 2V}b_0^{*2}\sum_p g(p) b(p) b({-p})
+{\lambda \over 2V}b_0^{2}\sum_p g(p) b^*(p) b^*({-p})
$$
This implies that the ground state of $\Gamma$ has the energy
$$
E_1={\lambda N_0^2\over 2V}g(0)-\mu N_0
$$

The chemical potential is defined by the equation
$$
{\partial E_1\over\partial N_0}=0
$$
and in our case it is equal to
$$
\mu={\lambda N_0\over V}g(0)
$$
With this chemical potential and making the canonical transformation
$b(p)\mapsto b(p)e^{i\phi}$ we cancel the phase of the complex number
$b_0=\sqrt{N_0}e^{i\phi}$ so that the Hamiltonian $\Gamma$ becomes
$$
\Gamma=-{\lambda N_0^2\over 2V}g(0)+
\sum_{p\ne 0}\left(\omega(p)+{\lambda N_0\over V}g(p) \right)b^*(p) b(p) +
{\lambda N_0\over 2V}\sum_{p\ne 0} g(p) (b^*(p) b^*({-p})+b({-p}) b({p}))
$$
Let us make a canonical transformation to diagonalize this quadratic
Hamiltonian. We consider the following transformation
$$
a(p)=u_pb(p)+v_pb^*({-p})
$$
$$
a^*(p)=u_pb^*(p)+v_pb({-p})
$$
$$
u_p^2-v_p^2 =1,\quad  u_{p}=u_{-p},v_{p}=v_{-p};\quad u_p,v_p\in {\bf R}
$$
The inverse transformation is given by
$$
b(p)=u_p a(p)-v_p a^*({-p})
$$
$$
b^*(p)=u_p a^*(p)-v_p a({-p})
$$

After this canonical transformation the off--diagonal terms in the Hamiltonian
(coefficients of $a^*({p})a^*({-p})$ and $a({-p})a({p})$)
are equal to
\begin{equation}\label{offdiag}
-u_pv_p\left(\omega(p)+{\lambda N_0\over V}g(p) \right)
+(u_p^2+v_p^2){\lambda N_0\over 2V}g(p)
\end{equation}
and the diagonal terms (coefficients of $a^*({p})a({p})$)
are equal to
\begin{equation}\label{diag}
(u_p^2+v_p^2)\left(\omega(p)+{\lambda N_0\over V}g(p) \right)
-4u_pv_p{\lambda N_0\over 2V}g(p)
\end{equation}
Since $u^2_p-v^2_p=1$ we can use the hyperbolic parametrization
$$
u_p=\hbox{ ch } x,\qquad v_p=\hbox{ sh } x
$$
The compensation equation (vanishing of off--diagonal terms
(\ref{offdiag})) in this parametrization takes the form
$$
\hbox{ th } 2x =
{{\lambda N_0\over V}g(p)\over
\omega(p)+{\lambda N_0\over V}g(p)}
$$
This implies
$$
u_p^2+v_p^2=\hbox{ ch } 2x=
{1\over\sqrt{1-
\left(
{{\lambda N_0\over V}g(p)\over
\omega(p)+{\lambda N_0\over V}g(p)}
\right)^2
}}
$$
$$
2u_pv_p=\hbox{ sh } 2x =
{1\over\sqrt{1-
\left(
{{\lambda N_0\over V}g(p)\over
\omega(p)+{\lambda N_0\over V}g(p)}
\right)^2
}}
{{\lambda N_0\over V}g(p)\over
\omega(p)+{\lambda N_0\over V}g(p)}
$$
Therefore for (\ref{diag}) we get
\begin{equation}\label{E(p)}
E(p)=\sqrt{\omega^2(p)+{2\lambda N_0\over V}\omega(p)g(p)}
\end{equation}
Condition $g(0)>0$ that provides positivity of the value
under the square root corresponds to the domination of repulsion,
cf. \cite{BTS}.

The Hamiltonian takes the form
\begin{equation}\label{Gamma}
\Gamma=-{\lambda N_0^2\over 2V}g(0)+\sum_{p\ne0}E(p)a^*(p)a(p)
\end{equation}

The state for hte new Hamiltonian (\ref{Gamma})
corresponds to a state of the non--ideal Bose gas modified due to
interaction.

Let us discuss the Bogoliubov--Landau condition
for Hamiltonian the (\ref{Gamma}). We get for (\ref{energydiff})
\be\label{LAforGamma}
E(p)+{m\left(u-{p\over m}\right)^2\over2}-{mu^2\over2}=E(p)-pu+{p^2\over2m}
\ee
Since $E(p)=c|p|$ for small $p$ we can get the situation when for $u<c$
the energy of excitation is positive and transition from
condensate to normal state proceeds with consumption of energy.
In this case we can apply the discussion of (\ref{example1})
which shows that the Bogoliubov--Landau condition is applicable
to (\ref{LAforGamma}).

\section{Condensate--normal phase interaction in the stochastic limit}

In the present section we will discuss how to describe
the Bogoliubov--Landau condition by the stochastic limit approach.

We will consider the stochastic limit of a system with phase
transition. We will see that the phase transition leads to
arising of non--linear master and kinetic equations.This shows that
phase transition and the stochastic limit procedure do not commute:
to investigate a system with phase transition by the stochastic limit
we should describe first the phase transition.

We introduce the Hamiltonian for condensate--normal state
interaction and investigate it in the stochastic limit.

Consider a system with two Bose fields (phases)
$$
[c(p),c^+(p')]=\delta(p-p'),~[a(k),a^+(k')]=\delta(k-k')
$$
We describe the condensate by the Bose field $c(p)$
(system degrees of freedom)
labeled by a velocity index $p$ with free Hamiltonian
$$
H_c=\int \varepsilon(p) c^*(p) c(p) d^3p
$$
with dispersion
$$\varepsilon(p)={mp^2\over 2}$$
and state
$$
\langle c^*(p) c(p')\rangle = N(p)\delta(p-p')
$$
This state describes the distribution of condensate over velocities.

The second Bose field $a(p)$ (reservoir degrees of freedom)
describes excitations of the non--ideal Bose gas
(normal phase)
which diagonalize the interaction (considered in the previous section).
This excitations
correspond to pairs of particles with opposite momenta and has free
Hamiltonian
$$
H_{ns}=\int E(p) a^*(p) a(p) d^3p
$$
The state of $a(p)$ is an equilibrium
$$
\langle a^*(k)a(k')\rangle=\delta(k-k'){1\over e^{\beta\omega(k)}-1}
$$

The total Hamiltonian will be
\begin{equation}\label{cxii}
H=H_c+H_{ns}+\lambda H_I
\end{equation}
with the interaction
\be\label{intccxi}
H_I=\int\int \overline{f(k,p)}  c^*(p)c(p-k)a(k)d^3pd^3k +\hbox{ h.c.}
\ee
where $f(u,p)$ is the form--factor (complex valued test function).

After this we apply to the Hamiltonian (\ref{cxii}) the stochastic limit
approach: by the stochastic golden rule we construct the master equation
for the density of particles in the condensate $c^*(p)c(p)$
and prove that the stochastic golden rule gives
the Bogoliubov--Landau condition for this system.

This approach is close to the discussion in Abrikosov--Gorkov--Dzialoshinskii
book:
{\it Consider the Bose liquid with zero temperature that flows with velocity
${\bf v}$. In the presence of friction in the liquid there
will be elementary excitations with different velocities.}
An analysis of these elementary excitations gives
the Bogoliubov--Landau condition,
cf \cite{Landau}, \cite{BB}, \cite{AGD}.

The friction is due to the interaction term (\ref{intccxi}).
The rescaled free evolution of the interaction (\ref{intccxi})
(of the collective field) is equal to
\be
{\cal A}_\lambda(p,k,t)={1\over\lambda}\,
e^{{itH_{0}\over\lambda^2}}c^+(p)a(k)c(p-k)
e^{-{itH_{0}\over\lambda^2}}=
{1\over\lambda}\,c^+(p)a(k)c(p-k)e^{-itE(p,k)
/\lambda^2}
\label{evolved}
\ee
\be
{\cal A}^+_\lambda(p,k,t)=
{1\over\lambda}\,e^{{itH_{0}\over\lambda^2}}
c^+(p-k)a^+(k)c(p)e^{-{itH_{0}\over\lambda^2}}=
{1\over\lambda}\,
c^+(p-k)a^+(k)c(p)e^{itE(p,k)/\lambda^2}
\label{evolved'}
\ee
where
\be
E(p,k)=E(k)+\varepsilon (p-k)-\varepsilon (p)
\label{en}
\ee
is the corresponding energy difference.

The difference of energies (\ref{en}) coincides with the
difference of energies in the Bogoliubov--Landau condition (\ref{LAforGamma})
that justify the fact that Hamiltonian (\ref{cxii}) describe the non--ideal
Bose gas considered above.

The stochastic limit of the rescaled evolution of the interaction
(\ref{evolved}) gives rise to a quantum white noise (master field):
$$
\lim_{\lambda\to 0}{\cal A}_\lambda(p,k,t)=B(p,k,t)
$$
After the stochastic limit we cannot separate in the master field
the degrees of freedom of reservoir and normal state.
This means that the degrees of freedom of the condensate
and of the normal phase become entangled even at kinematical level,
cf. \cite{entangled}.

The master field $B(p,k,t)$ will satisfy the variant of quantum Boltzmann
statistics given by the following theorem

\bigskip

\noindent
{\bf Theorem 1.}\quad {\sl
The entangled noise algebra is generated by
$B_1(p,k,t)$, $B_2(p,k,t)$, $n(p)$ where
$$
B(p,k,t)=B_1(p,k,t)+B_2^*(p,k,t)
$$
with relations
$$
B_1 B_2^*=B_2 B_1^*=0
$$
\be\label{skb1b1dag}
B_1(p,k,t) B_1^*(p',k',t')=2\pi\delta(t-t')\delta(E(p,k))n(p)
(N(p-k)+1)\delta(p-p')
{\delta(k-k')\over 1-e^{-\beta\omega(k)}}
\ee
\be\label{skb2b2dag}
B_2(p,k,t) B_2^*(p',k',t')=2\pi\delta(t-t')\delta(E(p,k))n(p-k)
(N(p)+1)\delta(p-p')
{\delta(k-k')\over e^{\beta\omega(k)}-1}
\ee
$$
[n(p'), B_1^{\mp}(p,k,t)]=(\pm)(\delta(p'-p)
-\delta(p'-p+ k)) B_1^{\mp}(p,k,t)
$$
$$
[n(p'), B_2^{\mp}(p,k,t)]=(\mp)(\delta(p'-p)
-\delta(p'-p+ k)) B_2^{\mp}(p,k,t)
$$
$$
[n(p),n(p')]=0
$$
Here $B_i^{-}=B_i$, $B_i^{+}=B_i^*$.
}
\bigskip

The zero temperature version of this theorem was obtained in \cite{entangled}.
The one particle sector relations ($n(p)=1$) were investigated in \cite{Kyoto},
\cite{hotfree}, \cite{qdeform}.

Consider now the evolution in the stochastic limit for the considered
Hamiltonian.
By the stochastic golden rule, cf. \cite{book} we get

\bigskip

\noindent{\bf Theorem 2.}{\sl\quad
The stochastic differential
equation for $U_t$ has the form
\begin{equation}
{dU_t}\,=\biggl(-i\int dkdp\left(\overline f(k,p)dB(p,k,t)+
dB^{\dag}(p,k,t)f(k,p)\right)-
$$
$$
-dt\,(f|f)_- -dt\,\overline{(f|f)}_+ \biggr)U_t
\end{equation}
where the stochastic diferentials satisfy the Ito table
\be
dB(p,k,t)dB^{\dag}(p',k',t)=
2\pi n(p)\delta(\omega(k)+\varepsilon(p-k)-\varepsilon(p))
$$
$$
(N(p-k)+1)\delta(p-p')
{\delta(k-k')\over 1-e^{-\beta\omega(k)}}dt
\ee
\be
dB^{\dag}(p,k,t)dB(p',k',t)=
2\pi n(p-k)\delta(\omega(k)+\varepsilon(p-k)-\varepsilon(p))
$$
$$
(N(p)+1)\delta(p-p')
{\delta(k-k')\over e^{\beta\omega(k)}-1}dt
\ee
$$
(f|f)_-=
\int dkdp |f(k,p)|^2
{-in(p)(N(p-k)+1)\over \omega(k)+\varepsilon(p-k)-\varepsilon(p)-i0}
{1\over 1-e^{-\beta\omega(k)}}
$$
$$
(f|f)_+=
\int dkdp |f(k,p)|^2
{-in(p-k)(N(p)+1)\over \omega(k)+\varepsilon(p-k)-\varepsilon(p)-i0}
{1\over e^{\beta\omega(k)}-1}
$$
}

One particle version of this quantum stochastic differential equation was
obtained and investigated in \cite{polaron}.

\section{Master equation and superfluidity}

In the present section we consider the master equation for
the stochastic dynamics described in the previous section.
The master equation is the equation for the expectation over the degrees
of freedom of the reservoir (normal state in the considered case)
of the Heisenberg dynamics for some observable $X$
\be\label{master}
{d\over dt}X_t={d\langle U^*_t X U_t \rangle\over dt}=
{d\over dt}\langle j_t(X)\rangle=\langle j_t(\theta_0(X))\rangle
\ee
with $U_t$ defined by theorem 2.

Consider the master equation for the density $n(p)=c^*(p)c(p)$.
The stochastic golden rule (application of theorems 2 and 1) gives
\be\label{mastersfluidity}
{d\over dt}n_t(q)=\int dkdp |f(k,p)|^2(\delta(q-p)-\delta(q-p+k))
2\pi\delta(E(k)+\varepsilon(p-k)-\varepsilon(p))
$$
$$
\biggl(
n_t(p-k)(N(p)+1)
{1\over e^{\beta E(k)}-1}
- n_t(p)(N(p-k)+1)
{1\over 1-e^{-\beta E(k)}}
\biggr)
\ee
Since $n_t(p)$ is in the abelian subalgebra this equation is classical
and we can consider $n(p)$  as a classical distribution.

Taking the integral over $p$ one gets
\be\label{mastersfluidity1}
{d\over dt}n_t(q)=2\pi\int dk\biggl(
|f(k,q|^2
\delta(E(k)+\varepsilon(q-k)-\varepsilon(q))
$$
$$
\biggl(
n_t(q-k)(N(q)+1)
{1\over e^{\beta E(k)}-1}
- n_t(q)(N(q-k)+1)
{1\over 1-e^{-\beta E(k)}}
\biggr)-
$$
$$
-|f(k,q+k)|^2
\delta(E(k)+\varepsilon(q)-\varepsilon(q+k))
$$
$$
\biggl(
n_t(q)(N(q+k)+1)
{1\over e^{\beta E(k)}-1}
- n_t(q+k)(N(q)+1)
{1\over 1-e^{-\beta E(k)}}
\biggr)\biggr)
\ee
Let us discuss the connection of the Bogoliubov--Landau condition
with (\ref{mastersfluidity1}).
Since the dispersion (\ref{E(p)}) for small $|k|$ is radiative
$E(k)=c|k|$ and $\varepsilon(p)={p^2\over 2m}$ we get
$$
E(k)+\varepsilon(q-k)-\varepsilon(q)=
c|k| - {1\over m}qk + {k^2\over 2m} >0, \quad {1\over m}|q|< c
$$
$$
E(k)+\varepsilon(q)-\varepsilon(q+k)=
c|k| - {1\over m}qk - {k^2\over 2m}
$$
This implies that the first $\delta$--function in (\ref{mastersfluidity1})
vanishes for ${1\over m}|q|< c$.
Let us note that the second $\delta$--function in (\ref{mastersfluidity1})
does not vanish for any $q$.

To get a superfluid
motion in (\ref{mastersfluidity1})
it is sufficient to consider such $n(q)$ that the following products
will vanish for arbitrary $q$ and $k$:
\be\label{prod1}
n(q)\delta(E(k)+\varepsilon(q-k)-\varepsilon(q))=0
\ee
\be\label{prod2}
n(q-k)\delta(E(k)+\varepsilon(q-k)-\varepsilon(q))=0
\ee
If $n(q)$ would satisfy (\ref{prod1}), (\ref{prod2})
then the RHS of (\ref{mastersfluidity1}) would vanish.

To satisfy (\ref{prod1})  it is sufficient to take the support
of $n(q)$ is concentrated in the ball $|q| \le mc$.
But (\ref{prod2}) can not be satisfied when the support
of $n(q)$ is concentrated in the ball $|q| \le mc$.
This implies that (\ref{mastersfluidity1})
does not describe a superfluid liquid and to investigate
superfluidity in the stochastic limit
one needs some additional arguments.

To overcome this problem we propose the following construction.

First, since $N(p)>>1$ we can identify $N$ and $N+1$.

Second (and most important), we identify $n_t(p)$ with
expectation $\langle n(p)\rangle=N(p)$ in (\ref{mastersfluidity1})
(we substitute $N(p)$ by $n_t(p)$ in this formula).
The identification
\be\label{identify}
\langle n(p)\rangle=N(p):=n_t(p)
\ee
means that we consider the stochastic limit with the state
depending on time in the slow time scale of the stochastic limit.
This modifies the stochastic limit
procedure in order to take into account the effects
of phase transition.

This condition may be justified by the following argument.
Equation (\ref{master}) for $X=n(q)$ is
\be\label{master1}
{d\over dt} \langle U^*_t n(q) U_t\rangle =
\langle j_t(\theta_0(n(q)))\rangle
\ee
where $\theta_0(n(q))$ is equal to the RHS of (\ref{mastersfluidity1}).
Then we apply the identification (\ref{identify}) for $t=0$
to (\ref{master1}). We get,
in the integral on RHS of (\ref{master1}),
combinations of terms of the folowing form
$$
j_t (n(q)n(q-k))=j_t(n(q))j_t(n(q-k))
$$
since $j_t$ is a homomorphism.
After this we apply the semiclassical approximation
$$
\langle j_t(n(q))j_t(n(q-k))\rangle =
\langle j_t(n(q))\rangle \langle j_t(n(q-k))\rangle=
n_t(q) n_t(q-k)
$$
that justifies (\ref{identify})  for any $t$.

Condition (\ref{identify}) introduces the self--interaction
into the considered model and makes equation
(\ref{mastersfluidity1}) nonlinear.
Condition (\ref{identify}) implies for (\ref{mastersfluidity1})
the following:
\be\label{mastersfluidity2}
{d\over dt}n_t(q)=-2\pi\int dk\biggl(
|f(k,q|^2
\delta(E(k)+\varepsilon(q-k)-\varepsilon(q)) n_t(q)n_t(q-k)-
$$
$$
-|f(k,q+k)|^2
\delta(E(k)+\varepsilon(q)-\varepsilon(q+k))n_t(q)n_t(q+k)
\biggr)
\ee
Let us note that this equation is nonlinear (quadratic).

For (\ref{mastersfluidity2}) the condition of superfluidity
is reduced to (\ref{prod1}) (condition (\ref{prod2}) can be ignored).
This implies that (\ref{mastersfluidity2}) describes a superfluid flow.
We get that in the stochastic limit the condition
\be\label{slow}
\hbox{ Supp }n(q)\subset\{q: |q| \le mc \}
\ee
is sufficient for superfluidity in the sence that,
under this condition, the right hand side of (\ref{mastersfluidity2})
is zero and the density $n_t(q)$, of the condensate, is constant.

We call equation (\ref{mastersfluidity2})  the equation
of motion of superfluid liquid.

Non--linear master equation (\ref{mastersfluidity2})
is an example of a general phenomenon.
Application of the stochastic limit to systems with phase transitions
will generally create non--linear master and kinetic equations.
Non--linearity will enter through the self--interaction
given by an analogue of (\ref{identify})  for a certain phase
(an analogue of the condensate considered in the present paper).

\section{Conclusion}

We summarize our main results as follows:

\medskip

\noindent
1) The Bogoliubov--Landau condition of superfluidity
follows from the stochastic limit approach.

\medskip

\noindent
2) The stochastic limit approach gives not only a condition of
superfluidity but also a natural candidate for
the equation of motion of the superfluid liquid (\ref{mastersfluidity2}).

\medskip

\noindent
3) This equation is nonlinear (quadratic).

\medskip

\noindent
4) Without the introduction of a non--linearity (identification of
$n(p)$ and $N(p)$) it is impossible to get superfluidity
in the stochastic limit approach.

\medskip

\noindent
5) Nonlinear master equations should be a general feature of the stochastic
limit of self--interacting systems.

\bigskip

\centerline{\bf Acknowledgements}
The authors are grateful to
I.V.Volovich and H.-D.Doebner for discussions.
Sergei Kozyrev is grateful to Centro Vito Volterra and Luigi Accardi
for kind hospitality. This
work was partially supported by INTAS 9900545 grant.
Sergei Kozyrev was partially supported by RFFI 990100866
grant.

\end{document}